\documentstyle[preprint,aps,eqsecnum,tighten,twoside]{revtex}

\def\be{\nopagebreak[3]\begin{equation}}
\def\ee{\end{equation}}
\def\ba{\nopagebreak[3]\begin{eqnarray}}
\def\ea{\end{eqnarray}}

\def\I{\it I}
\def\d{\partial}
\def\={\cong}

\begin{document}
\title{Behavior of Einstein-Rosen Waves at null infinity}
\author{Abhay Ashtekar${}^{1}$, Ji\v r{\'\i} Bi\v c\'ak${}^{2}$,
and Bernd G. Schmidt${}^{3}$}
\address{${}^1$ Center for Gravitational Physics and Geometry\\
Department of Physics, Penn State, University Park, PA 16802, USA}
\address{${}^2$ Department of Theoretical Physics, Charles University\\
V Hole\v sovi\v ck\'ach 2, 180 00 Prague 8, Czech Republic}
\address{${}^3$ Max-Planck-Institut f{\"u}r Gravitationsphysik,\\
Schlaatzweg 1, 14473 Potsdam, Germany}
\maketitle
\pagenumbering{arabic}

\begin{abstract}
The asymptotic behavior of Einstein-Rosen waves at null infinity in 4
dimensions is investigated in {\it all} directions by exploiting the
relation between the 4-dimensional space-time and the 3-dimensional
symmetry reduction thereof. Somewhat surprisingly, the behavior in a
generic direction is {\it better} than that in directions orthogonal
to the symmetry axis. The geometric origin of this difference can be
understood most clearly from the 3-dimensional perspective.
\end{abstract}

\begin{section} {Introduction}
\label{s1}

Although the literature of Einstein-Rosen waves is quite rich (see,
e.g., the references listed in the companion paper \cite{[1]}) it
appears that there is only one article that discusses the asymptotic
behavior of these waves at infinity in 4-dimensions: the paper by
Stachel \cite{[2]} written already in the sixties. Moreover, even in
this work, Stachel deals solely with the directions {\it orthogonal}
to the axis of symmetry, i.e., to the $\d/\d z$ Killing field. The
purpose of this note is to analyze the asymptotic structure in {\it
all} directions.

Since these space-times admit a translational Killing field, one would
expect them not to be asymptotically flat. This is precisely what
Stachel discovered in directions orthogonal to the symmetry axis.
Somewhat surprisingly, however, we will find that the fall-off is much
better in generic directions. Indeed, if one restricts oneself to the
``time-symmetric'' situation, one finds that, in all other directions,
curvature peels normally and a regular null infinity, $\I$, exists. In
fact, all radiation is concentrated along the two generators in which
the null geodesics orthogonal to the symmetry axis meet $\I$. In other
directions, there is curvature but no radiation. If one goes beyond
the ``time-symmetric'' case, the behavior is not as nice; $\I$ may
have a logarithmic character \cite{[3],[4]}.  That is, the metric does
admit Bondi-type expansions but in terms of $r^{-j} \ln^i
r$. Nonetheless, even this behavior is better than the one encountered
in the directions orthogonal to the symmetry axis.

The key idea behind our analysis is to exploit the relation between
4-dimensional Einstein-Rosen waves and the associated 3-dimensional
geometry on the manifold of orbits of the translational Killing field.
Since the translational Killing field has been ``factored out'' in the
passage to 3 dimensions, the 3-dimensional space-time is
asymptotically flat at null infinity \cite{[1]} and, as we will see,
also admits a regular time-like infinity. To analyze the behavior of
the 4-dimensional metric, we can draw on this 3-dimensional
information. We will find that the behavior at $\I$ in generic
directions in 4 dimensions is dictated by the behavior of various
fields at {\it time-like} infinity in 3 dimensions. Since this
3-dimensional time-like infinity is regular, the behavior in generic
directions in 4 dimensions is better than what one might naively
expect.

The plan of the paper is as follows. In Sec. II, we shall present the
3-dimensional structure. In Sec. III, we will use this structure to
investigate 4-dimensional null infinity. Appendix A spells out the
relation between 3- and 4-dimensional curvatures.
\end{section}

\begin{section}{3-Dimensional description}
\label{s2}

This section is divided into three parts. In the first, we briefly
recall the symmetry reduction procedure and apply it to obtain the
3-dimensional equations governing Einstein-Rosen waves. (For details,
see \cite{[1]}.)  This procedure reduces the task of finding a
4-dimensional Einstein-Rosen wave to that of finding a solution to the
wave equation on 3-dimensional {\it Minkowski} space.  In the second
part, we analyze the asymptotic behavior of these solutions to the
wave equation at time-like infinity of the 3-dimensional Minkowski
space.  In the third part, we combine the results of the first two to
analyze the asymptotic behavior of the 3-dimensional metric associated
with Einstein-Rosen waves at its time-like infinity. We show that this
time-like infinity is regular. Although this result is not needed
directly for our main result, it is included because it complements
the general analysis of 3-dimensional null infinity presented in
\cite{[1]}.

\begin{subsection} {Symmetry reduction}

Recall first that the metric of a vacuum space-time with two
commuting, hypersurface orthogonal space-like Killing vectors can
always be written locally as \cite{[5]}
\be ds^2=e^{2\psi}dz^2+ e^{2(\gamma
-\psi)}(-dt^2+d\rho^2)+\rho^2e^{-2\psi}d\phi^2\ ,
\label{(2.1)}
\ee
where $\rho$ and $t$ (the ``Weyl canonical coordinates'') are defined
invariantly and $\psi=\psi(t,\rho)$, $\gamma=\gamma (t,\rho)$.  (Here,
some of the field equations have been used.) Einstein-Rosen waves have
cylindrical symmetry; the Killing field $\d/\d z$ is translational and
$\d/\d \phi$ is rotational and keeps a time-like axis fixed. Then the
coordinates used in (2.1) are unique up to a translation $t\mapsto
t+a$.

The 3-manifold is obtained by quotienting the 4-dimensional space-time
by the orbits of the $\d/\d z$ Killing field and is thus
co-ordinatized by $t, \rho$ and $\phi$. The 4-metric naturally induces
a 3-metric $d\bar\sigma^2$ on this manifold and the 4-dimensional
Einstein's equations can be expressed on the 3-manifold as a system of
coupled equations involving the induced 3-metric and the norm of the
Killing field $\d/\d z$, which, from the 3-dimensional perspective can
be regarded as a (scalar) matter field.  It is well-known (see
\cite{[1],[6],[16]}), however, that the field equations simplify
considerably if we rescale the induced 3-metric $d\bar\sigma^2$ by
$\exp\,(2\psi)$, the square of the norm of the Killing field, i.e., in
terms of the 3-metric
\be
d\sigma^2=g_{ab}dx^adx^b=e^{2\gamma}(-dt^2+d\rho^2)
+\rho^2d\phi^2\ .
\label{(2.2)} \ee
The 4-dimensional vacuum equations are then equivalent to the set
(cf. Eqs. (2.12)--(2.15) in the preceding paper \cite{[1]}):
\ba
\gamma''-\ddot\gamma +\rho^{-1}\gamma' &=& 2\dot\psi^2\ , \\
-\gamma''+\ddot\gamma +\rho^{-1}\gamma' &=& 2\psi'^2\ , \\
\rho^{-1}\dot\gamma &=& 2\dot\psi\psi'\ ,
\label{(2.5)}\ea
and
\be -\ddot\psi+\psi''+ \rho^{-1}\psi'=0 \ , \label{(2.6)} \ee
on the 3-manifold, where the dot and the prime denote derivatives with
respect to $t$ and $\rho$ respectively.  The last equation is the wave
equation for the non-flat 3-metric (2.2) {\it as well as for the flat
metric obtained by setting} $\gamma=0$.  This is a key simplification
for it implies that the equation satisfied by the matter source $\psi$
decouples from the equations (2.3)--(2.5) satisfied by the
metric. Furthermore, these latter equations reduce simply to:
\be \gamma ' =\rho\,(\dot\psi^2+\psi'^2)\ , \label{(2.7)}\ee
\be \dot\gamma = 2\rho\dot\psi\psi '\ . \label{(2.8)}\ee
Thus, we can first solve for the axi-symmetric wave equation (2.6) for
$\psi$ {\it on Minkowski space} and then solve (2.7) and (2.8) for
$\gamma$ --the only unknown metric coefficient-- by quadratures.
(Note that (2.7) and (2.8) are compatible because their integrability
condition is precisely (2.6).)
\end{subsection} 
\goodbreak

\begin{subsection}{Asymptotic behavior of scalar waves}

In this subsection we will focus on the axi-symmetric wave equation in
3-dimensional Minkowski space and analyze the behavior of its
solutions $\psi$ near time-like infinity of Minkowski space. (For
behavior at null infinity, see \cite{[1]}.)

We begin with an observation.  The ``method of descent'' from the
Kirchhoff formula in 4 dimensions gives the following representation
of the solution of the wave equation in 3 dimensions, in terms of
Cauchy data $\Psi_0=\psi(t=0,x,y), \Psi_1=\psi_{,t}(t=0,x,y)$:
\ba
\psi(t,x,y) &=& {1\over 2 \pi}\ {\partial\over \partial t}
\int\!\!\!\int_{S(t)}
{\Psi_0(x',y')dx'dy' \over [t^2-(x-x')^2-(y-y')^2]^{1/2}}
\nonumber\\
&+&{1\over 2 \pi} \int\!\!\!\int_{S(t)}
{\Psi_1(x',y')dx'dy' \over [t^2-(x-x')^2-(y-y')^2]^{1/2}}\ ,
\label{(2.9)}\ea
where $S$ is the disk 
$$ (x-x')^2+(y-y')^2\le t^2 $$
in the initial Cauchy surface (see e.g. \cite{[7]}).  We will assume
that the Cauchy data are axially symmetric and of compact support.

In the preceding paper \cite{[1]} (see Eq. (2.23)) we have shown that
on each null hypersurface $u=t-\rho = const$ the solution (2.9) can be
expanded in the form
\be
\psi(u,\rho)={1\over\sqrt\rho}\left(f_0(u)+
\sum_{k=1}^\infty{f_k(u)\over\rho^k}\right)\ ,
\label{(2.10)}\ee 
where the coefficients in this expansion are determined by integrals
over the Cauchy data. This is the behavior of $\psi$ at null infinity
$I$.

Let us now investigate the behavior of the solution (2.9) near
time-like infinity $i^+$ of the 3-dimensional Minkowski space.
Setting
\be t=U+\kappa\rho\ , \ \ \ \kappa>1\ ,\label{(2.11)}\ee
we wish to find $\psi$ for $\rho\to\infty$ with $U$ and $\kappa$
fixed.  For large enough $\rho$ the region of integration is contained
in the cone. Hence we have to perform the derivative in (2.9) only in
the integrand. We obtain
\ba
&&2\pi\psi(t,\rho)=\nonumber\\
&&-{\kappa\rho+U\over [(\kappa^2-1)\rho^2]^{3/2}}
\int_0^\infty\!\!\int_0^{2\pi}\!\!\Psi_0\rho'd\rho' d\phi'
\left[1+{2(\kappa U+\rho'\cos\phi')\over \kappa^2-1}\ {1\over\rho}
+{U^2-\rho'^2\over \kappa^2-1}\ {1\over\rho^2}
\right]^{-3/2} \nonumber\\
&&+{1\over [(\kappa^2-1)\rho^2]^{1/2}}
\int_0^\infty\!\!\int_0^{2\pi}\!\!\Psi_1\rho'd\rho'd\phi'
\left[1+{2(\kappa U+\rho'\cos\phi')\over \kappa^2-1}\ {1\over\rho}
+{U^2-\rho'^2\over \kappa^2-1}\ {1\over\rho^2}
\right]^{-1/2}. \nonumber\\
\label{(2.12)}\ea
The integrand can again be expanded in $\rho^{-1}$ (or $t^{-1}$) but
the leading term is $\rho^{-1}$. By contrast, at {\it null} infinity
of the 3-dimensional space-time, $\psi$ falls-off only as
$\rho^{-1/2}$\-  (see Eq. (2.10) and \cite{[1]} for details). We will
see that it is this difference that makes the behavior of the 4-metric
along generic directions better than that along directions orthogonal
to the symmetry axis.
 
The explicit expressions of the first few terms in the expansion of
$\psi$ is given by:
\ba
\psi&=& {L\over(\kappa^2-1)^{3/2}}\left[ -{\kappa\over\rho^2}+
{(2\kappa^2+1)U\over\kappa^2-1}\ {1\over\rho^3} +O(\rho^{-4})\ \right]
\nonumber\\
&+&{J\over(\kappa^2-1)^{1/2}}
\left[ {1\over\rho}-{\kappa U\over\kappa^2-1}\ {1\over\rho^2}
+O(\rho^{-3})\ \right]\, , 
\label{(2.13)}\ea
where
\be L=\int_0^\infty\Psi_0(\rho')\rho'd\rho'\ \ ,\ \
J=\int_0^\infty\Psi_1(\rho')\rho'd\rho'\ \ \ \  .
\label{(2.14)}\ee
By expressing $\rho$ in terms of $t$ using (2.11), we may rewrite
(2.13) as a series in $t^{-1}$:
\ba
\psi&=& {L\over(\kappa^2-1)^{3/2}}
\left[ -{\kappa^3\over t^2}+{3\kappa^3 U\over\kappa^2-1}\ 
{1\over t^3} +O(t^{-4})\ \right] \nonumber\\
&+&{J\over(\kappa^2-1)^{1/2}} \left[ 
{\kappa\over t}-{\kappa U\over\kappa^2-1}\ {1\over t^2}
+O(t^{-3})\, \right]. \label{(2.15)}\ea  
The last formula is meaningful also for $\rho=0$ in the limit
$\kappa\to\infty$:
\be \psi= {-L\over t^2} + {J\over t} + O(t^{-3})\ . 
\label{(2.16)}\ee
The same result can be obtained from (2.9) directly. This concludes
our discussion of the asymptotic behavior of $\psi$ near time-like
infinity $i^\pm$.

We will conclude this sub-section with three remarks.

First, the explicit representation (2.9) of the solution in terms of
Cauchy data allows us to make the interesting observation that the
solution is actually {\it analytic} in its space-time dependence for
all points for which the data are within the past null cone.  To show
that all solutions with data of compact support are also analytic in a
neighborhood of future time-like infinity ${i}^+$ we have to use
conformal rescaling techniques. Let
\be d\sigma^2=-dt^2+dx^2+dy^2 \label{(2.17)}\ee
be the metric of 3-dimensional Minkowski space. The conformal factor
\be \Omega=(t^2-x^2-y^2)^{-1}\label{(2.18)}\ee
defines, by the rescaling ${d\tilde\sigma^2= \Omega^2 d\sigma^2}$,
again a flat space-time
\be  d\tilde\sigma^2=\Omega^2 d\sigma^2=-d\tilde t^2+d
\tilde x^2+d\tilde y^2\ , \label{(2.19)}\ee
where the coordinates $\tilde t,\tilde x,\tilde y$ are defined by the
relations (inversion)
\be \tilde t={t\over  t^2-x^2- y^2}\ ,\ 
\tilde x={x\over t^2-x^2-y^2}\ ,\  
\tilde y={y\over t^2-x^2- y^2}\ .  \label{(2.20)}\ee
The 3-dimensional scalar wave equation has the following behavior
under this conformal re-scaling:
\be \nabla^2\psi=0 \Longrightarrow
\tilde\nabla^2\tilde\psi=0 \ ,\ \ \  \tilde\psi=\Omega^{-1/2}\psi\ .
\label{(2.21)}\ee

{}From the above consideration we know that a solution $\psi$ with
data of compact support is analytic for points within and on the
future light cone of the point $\tilde t=a,\tilde x=\tilde y=0$, where
the value of $a$ is dictated by the support of the data.  Moreover the
series (2.10) is also analytic in $v=u+2\rho$ because of the
converging expansion in $\rho^{-1}$.  Hence after the inversion we
have a solution $\tilde\psi$ which is analytic on the extended null
cone.  Therefore it is analytic in a domain which includes a
neighborhood of $i^+$.

The second remark concerns the asymptotic behavior of $\psi$, regarded
as a solution to the wave equation {\it in 4 dimensions}. More
precisely, let us set
\be F(t,x,y,z)=\psi(t,x,y)\ ; \label{(2.22)}\ee
$F$ is independent of $z$. How does this solution behave at null
infinity of 4-dimensional Minkowski space?  The null geodesics in a
hypersurface $z=const$ are also null geodesics in 4-space and $F=\psi$
along these curves. Now, a solution of the 4-dimensional wave equation
is well behaved at null infinity if it falls off as $r^{-1}$ (where
$r$ is the standard radial coordinate).  Since the field $\psi$ falls
off only as $\rho^{-{1\over 2}}$ at null infinity in 3-dimensions
\cite{[1]}, the solution $F$ fails to define a finite radiation field
at null infinity in these directions.  For null lines {\it not}
contained in $z=const$ surfaces, on the other hand, the situation is
entirely different. Because such null lines project onto {\it
time-like} lines in $z=const$, the fall-off behavior is {\it much
better} and from (2.15) we obtain the $r^{-1}$ decay, necessary for
the radiation field to exist. Thus, in terms of a 4-dimensional
conformal rescaling, the rescaled field of $F$ will be well-defined on
4-dimensional null infinity {\it except for} the two null generators
determined by the ${\partial/\partial z}$-Killing vector.  We will see
in Section III that this behavior is the key to the understanding of
the asymptotics of 4-dimensional axi-symmetric space-times with a
further $\partial/\partial z$-Killing vector.

Finally, we wish to point out that the main results obtained in this
section continue to hold also for general data of compact support
which are not necessarily axi-symmetric. In particular, asymptotic
forms like (2.13) and (2.15) hold where, however, the coefficients
depend on $\phi$.  The assumption of compact support can also be
weakened to allow data which decay near spatial infinity sufficiently
rapidly so that we still obtain solutions smooth at null and time-like
infinities. This is the case, for example, with the
Weber--Wheeler--Bonnor pulse discussed in the following section.
\end{subsection}
\goodbreak

\begin{subsection}{Asymptotic behavior of the metric}

We now combine the results of the previous two sub-sections. Recall
from Eq. (2.2) that the 3-dimensional metric $d\sigma^2$ has a single
unknown coefficient, $\gamma(t, \rho)$, which is determined by the
solution $\psi(t, \rho)$ to the wave equation in Minkowski space
(obtained simply by setting $\gamma= 0$).  The asymptotic behavior of
$\psi(t,\rho)$ therefore determines that of the 3-metric.

At null infinity $I$, the asymptotic behavior (2.10) of $\psi$
implies that $\gamma$ has the form (see Eq.(2.32) in \cite{[1]}): 
\be
\gamma = \gamma_0 + \int_{-\infty}^u\left( -2(\dot f_0(u))^2+
\sum_{k=1}^\infty{g_k(u)\over\rho^k}\right)du\ .
\label{(2.23)}\ee
We now wish to determine the metric at $i^+$.  In the last sub-section
we found the asymptotic form of $\psi$ at ${i}^+$, more specifically,
at $\rho\to\infty$ (or $t\to\infty$ with $U=t-\kappa\rho$, $\kappa>1$,
fixed (see (2.11) --(2.16)). In order to get the asymptotic forms of
$\gamma$, we first express the field equations (2.7) and (2.8) for
$\gamma$ in terms of $U$ and $\rho$:
\be \gamma_{,U}=2\rho\psi_{,U}(\psi_{,\rho}-\kappa\psi_{,U})\ ,  
\label{(2.24)}\ee 
\be \gamma_{,\rho}=\rho\left[\psi_{,\rho}^2+(1-\kappa^2)
\psi_{,U}^2\right]\ .  \label{(2.25)}\ee
Substituting for $\psi$ from (2.13) and (2.14), and integrating
(2.24), (2.25), we obtain
\ba
\gamma &=&{L^2\over {4\ (\kappa^2-1)^4}}
         \left[ {{8\kappa^2+1}\over \rho^4} -
          {{24\kappa(1+2\kappa^2)}\over{\kappa^2-1}}\ 
        {U\over\rho^5}+O(\rho^{-6}) \ \right]\nonumber \\
&+&{J^2\over {2\ (\kappa^2-1)^2}}\left[
         {1\over\rho^2}-{{4\kappa}\over{\kappa^2-1}}\ 
         {U\over\rho^3}+O(\rho^{-4})
         \  \right]\, . \label{(2.26)}\ea
Note that we set the integration constant equal to zero. This is
because we can go to ${i}^+$ along the center $\rho=0$. More
precisely, since we required the regularity of the solution at
$\rho=0$, we have to set $\gamma=0$ there and, as a consequence of the
field equations for $\gamma$ in $(t,\rho)$-coordinates (cf. (2.8)),
$\gamma$ at $\rho=0$ cannot change with time.

By techniques developed e.g. in \cite{[5]} it can be now shown that
the space-time has a smooth time-like infinity. The analyticity of
$\psi$ at time-like infinity shown in the last sub-section and the
field equations imply that the metric ``rescaled by inversion'' is
analytic at ${i}^+$. In what follows, however, we will not use this
result; the fall-off properties (2.15) and (2.26) of $\psi$ and
$\gamma$ will suffice.
\end{subsection}
\end{section}
\goodbreak

\begin{section}{Null infinity in 4 dimensions}
\label{s3}

We can now return to the 4-metric (2.1) and analyze its behavior at null
infinity.  In the main part of this section, we will consider those
Einstein-Rosen waves for which the Cauchy data for $\psi$ in the
3-dimensional picture are (smooth and) of compact support. In the
4-dimensional picture, these solutions correspond to {\it pulses} of
Einstein-Rosen waves.
\goodbreak

\begin{subsection}{Formulation of the problem}

Let us begin by summarizing the behavior in the directions
perpendicular to the axis of symmetry. In these directions, the
fall-off of $\psi$ is the same as in our 3-dimensional treatment of
null infinity (see Eq. (2.23) in \cite{[1]} or Eq. (2.10)). However,
from the 4-dimensional perspective, $\psi$ is not a matter field but a
metric coefficient (see (2.1)) and the $1/\sqrt{\rho}$ fall-off of
$\psi$ is too slow for null infinity to exist in the sense of Penrose
\cite{[8]}. What is the situation with respect to curvature? In
Appendix A, we use the 3-dimensional results to compute the
4-dimensional Riemann tensor for these space-times. We find that, in
null directions perpendicular to the $\d/\d z$ Killing field, the
tensor decays only as $1/\sqrt{\rho}$, the behavior that Stachel first
discovered in his direct 4-dimensional treatment \cite{[2]}. (See the
complex components of the Riemann tensor with respect to the null
tetrad given by his Eqs.  (A4)--(A6), or just ${}^{(4)}R_{3030}$,
given in our Eq.(A5)).  As one would suspect from the behavior of the
metric coefficients, the curvature does not peel properly in these
null directions. Thus, although we have asymptotic flatness at null
infinity of the {\it 3-dimensional} space-time \cite{[1]}, the
4-metric fails to be asymptotically flat in null directions
perpendicular to the axis of rotation (i.e., along 4-dimensional null
lines whose projections approach null infinity in 3 dimensions).

In the rest of the section, we will discuss the fall-off in the {\it
remaining} null directions. We will find that, contrary to what one
might have expected at first, the asymptotic behavior is much {\it
better}. For time-symmetric initial data, the fall-off in fact
satisfies the Bondi-Penrose \cite{[8],[9]} conditions and null infinity
is smooth in these directions. Even for a generic data, null infinity
exists but may have a ``logarithmic behavior''; the conformally
rescaled metric is continuous but need not be differentiable
there. Note that even this behavior is better than the one in
directions orthogonal to the symmetry axis.  The reason, in a
nutshell, is that the fall-off of various fields along a generic null
direction in 4 dimensions is dictated by the fall-off of that field
along a {\it time-like} direction in the 3-dimensional treatment and,
as we saw in Sec. II, fields decay more rapidly at the 3-dimensional
time-like infinity than at the 3-dimensional null infinity.

To see this point in detail, let us begin with the Einstein-Rosen
metric (cf. (2.1)), 
\be  ds^2=e^{2\psi}dz^2+ e^{2(\gamma -\psi)}(-dt^2+d\rho^2)+
\rho^2e^{-2\psi}d\phi^2\, ,\label{(3.1)}\ee
where $\psi=\psi(t,\rho), \gamma=\gamma(t,\rho)$. If we pass from
coordinates ($\rho,z,\phi$) to spherical coordinates ($r, \theta,
\phi$), so that $\rho =r\sin\theta, z=r\cos\theta, \phi=\phi$, and
introduce flat--space retarded time $U=t-r$, we obtain (3.1) in the
form
\ba 
ds^2&=&-e^{2{(\gamma-\psi)}}dU^2-2 e^{2(\gamma-\psi)}dUdr 
+\left[e^{2\psi}-e^{2(\gamma-\psi)}\right]\cos^2\theta\ dr^2
\nonumber\\ 
&+&\left[e^{2(\gamma-\psi)}\cos^2\theta
+e^{2\psi}\sin^2\theta\right] r^2d\theta^2 \nonumber\\
&+&\left[e^{2(\gamma-\psi)}-e^{2\psi}\right]2r\sin\theta \cos\theta dr
d\theta +r^2\sin^2\theta e^{-2\psi}d\phi^2\, .
\label{(3.2)}\ea

Since we are considering waves with initial data of compact support in
the $(\rho,\phi)$-plane, we can use the results of Section II
directly. Recall that one approaches $i^+$ in 3 dimensions, fixing
$U=t-\kappa\rho, \kappa=const>1$ (cf. (2.11)). In the 4-dimensional
picture, this corresponds precisely to approaching the {\it null
infinity} of the flat metric defined by $t,r,\theta, \phi$
coordinates, along $\theta=const$, $\phi=const$, $U=t-r=const$, if we
set $\kappa={1/\sin\theta}$. The expansions of $\psi$ and $\gamma$,
corresponding to (2.13) and (2.26), thus have the following forms:
\ba
\psi&=&{L\over {\cos^3\theta}}\left[-{1\over {r^2}} + 
{({2+\sin^2\theta)U}\over{\cos^2\theta}}\
{1\over{r^3}}+O({1\over{r^4}}) \right] \nonumber\\
&+&\ {J\over {\cos\theta}}\left[{1\over r}-{U\over{\cos^2\theta}}\ 
{1\over{r^2}}+O({1\over{r^3}}) \right]\, , 
\label{(3.3)}\ea 
and
\ba
\gamma&=&{1\over 4}{{L^2\sin^2\theta}\over{\cos^8\theta}}
\left[(8+\sin^2\theta)\ {1\over{r^4}}
 -{{24\ (2+\sin^2\theta)U}\over{\cos^2\theta}}\ 
{1\over{r^5}}+O({1\over{r^6}}) \right]\nonumber\\
&+&{1\over 2}{{J^2\sin^2\theta}\over{\cos^4\theta}}\left[{1\over{r^2}}
-{{4U}\over{\cos^2\theta}}\ {1\over{r^3}}+O({1\over{r^4}}) 
\right]\ , \label{(3.4)}\ea
provided we stay away from $\theta={\pi\over 2}$, i.e.  directions
perpendicular to the axis. Our task now is to cast the 4-metric in a
Bondi form and show that the metric coefficients have the standard
fall-off.

We will carry out this task in the next two sub-sections.  We will
consider the cases $J=0, L\ne 0$, and $J\ne0, L= 0$ separately; since
we are interested only in the leading terms, the cross-terms $\sim LJ$
are not relevant. The expansions (3.3), (3.4) show that the fall--off
of $\psi$ is slower than that of $\gamma$. Hence, in the construction
of the Bondi system, we can focus primarily on $\psi$.
\end{subsection}
\goodbreak

\begin{subsection}{Time-symmetric case; $ J= 0$}

$L$ and $J$ are determined by the Cauchy data for scalar waves, and
the case $J=0, L\ne 0$ corresponds to time--symmetric data.  Keeping
just the first term in $\psi$ in the expansion (3.3) and substituting
into (3.2), we find the asymptotic form of the metric to read:
\ba
ds^2 &=-& \left[1+{{2L\over{\cos^3\theta}}}\ {1\over{r^2}}+
\dots\right]dU^2 -2\left[1+{{2L\over{\cos^3\theta}}}\ 
{1\over{r^2}}+\dots\right]dUdr \nonumber\\
&-&\left[{{4L\over{\cos\theta}}}\ {1\over{r^2}}+\dots\right] dr^2 
+\left[1+{{2L\over{\cos^3\theta}}}(\cos^2\theta-\sin^2\theta)\ 
{1\over{r^2}}+\dots \right]r^2d\theta^2 \nonumber\\
&+&\left[{{8L\over{\cos^2\theta}}}+\dots \right]\sin\theta dr d\theta
+\left[1+{{2L\over{\cos^3\theta}}}\ {1\over{r^2}}+\dots\right]
r^2\sin^2\theta d\phi^2\, .
\label{(3.5)}\ea 

In order that the metric be of Bondi's form, we will use the method
developed in \cite{[10]} to analyze space-times with a boost-rotation
symmetry. What we need is a coordinate system $\bar U,\bar r,
\bar\theta, \bar\phi=\phi$ such that
$$ g_{\bar U\bar U}= 1+O(\bar r^{-1}), \,\, \, g_{\bar U\bar r}= 1 
+O(\bar r^{-1})\ , $$
\be  g_{\bar U\bar\theta}= O(1), \,\,\,
g_{\bar\theta\bar\theta}= \bar r^2 + O(\bar r), \, \,\, 
g_{\bar\phi\bar\phi}=\bar r^2 \sin^2\bar\theta\, ,
\label{(3.6)}\ee
and, to all orders, 
\be g_{\bar r\bar r}= g_{\bar r\bar\theta}=0,\
g_{\bar\theta\bar\theta}g_{\bar\phi\bar\phi}=\bar r^4
\sin^2\bar\theta.\label{(3.7)}\ee 

Let us suppose the transformation leading to this form may be expanded
in powers of $\bar r^{-1}$: 
\ba
U &=& \pi^0(\bar U,\bar\theta) + \pi^1(\bar U,\bar\theta)\bar
r^{-1}+\pi^2(\bar U,\bar\theta)\bar r^{-2}+\dots,\nonumber\\
r&=&q(\bar U,\bar\theta)\bar r +\sigma^0(\bar U,\bar\theta) +
\sigma^1(\bar U,\bar\theta) \bar r^{-1} +\dots, \nonumber\\
\theta&=&\tau^0(\bar U,\bar\theta)+\tau^1(\bar U,\bar\theta)
\bar r^{-1 }+\tau^2(\bar U,\bar\theta)\bar r^{-2 }+\dots \, . 
\label{(3.8)}\ea

The requirements (3.6), (3.7) restrict the undetermined functions
$\pi, q, \sigma, \tau$. {}From the leading terms of $g_{\bar U\bar
U}$, $g_{\bar U\bar\theta}$ we first find that $q_{,\bar
U}=\tau^0_{,\bar U}=0$. The required form of
$g_{\bar\theta\bar\theta}$ and $g_{\bar\phi\bar\phi}$ in the leading
terms implies $(q)^2\tau^0_{,\bar\theta}=1$,
$(q)^2={\sin^2\theta}/{\sin^2\tau^0}$.  This can be solved for $q$ and
$\tau^0$ explicitly; however, further we assume
$q=\tau^0_{,\bar\theta}=1$ since the other choices just correspond to
coordinate systems connected by boosts along the symmetry axis
\cite{[9],[10]}. Then the requirement on the leading order term in
$g_{\bar U\bar r}$ implies that also $\pi^0_{,\bar U}=1$. The
fall--off conditions (3.6) are thus satisfied.

The conditions $g_{\bar\theta\bar\theta}g_{\bar\phi\bar\phi}=\bar r^4
sin^2\bar\theta + O(\bar r^{2})$ and $g_{\bar r\bar\theta}=O(\bar
r^{-1})$ lead to $\sigma^0=\tau^1=0$. It remains only to satisfy the
requirements (3.7).

The conditions $g_{\bar r\bar r}=0$ (to order $O(\bar r^{-2})$),
$g_{\bar r \bar\theta}=0$ (to $O(\bar r^{-1})$) and
$g_{\bar\theta\bar\theta}g_{\bar\phi\bar\phi}=\bar r^4
\sin^2\bar\theta$ (to $O(\bar r^{2})$) determine the functions
$\pi^1$, $\pi^2, \dots$, $\tau^2, \tau^3\dots,$ and $\sigma^1,
\sigma^2,\dots $. More specifically,  the vanishing of $g_{\bar r 
\bar r}$ to $\sim \bar r^{-2}$ implies $\pi^1{=2L}/{\cos\bar\theta}$, 
$g_{\bar r\bar\theta}=0$ to $\sim \bar r^{-1}$ leads to $\tau^2= -(1/2) 
\pi^1_{,\bar\theta}+2L{\sin\bar\theta}/{\cos^2\bar\theta}$, and
$g_{\bar\theta\bar\theta}g_{\bar\phi\bar\phi}=0$ in order $\bar r^2$
gives $\sigma^1=-{(L/{\cos\bar\theta}}$ ${+(1/2) \tau^2_{,\bar\theta})}
/\sin\bar\theta - (1/2)\tau^2\cot \bar\theta$. To determine the higher
order functions $\pi, \tau$ and $\sigma$, we have, of course, to
consider also the function $\gamma$ in the metric (3.2).  Calculations
then become lengthy. Nonetheless, they can be performed and one can
thus demonstrate the existence of the Bondi expansion for
time-symmetric waves. This establishes the existence of a smooth null
infinity in all directions except those perpendicular to the axis of
the symmetry.

Now, in axi-symmetric space-times, when a space-like Killing field
with circular orbits exists, there is a reduction of the asymptotic
symmetry group even if a ``global'' $\I$ does not exist, i.e., even if
$\I$ does not admit spherical cross-sections. Furthermore, in this
case, the Bondi news function has a local meaning \cite{[11]}. One can
therefore try to find it in the present case. In Bondi's coordinates
the news function is given by $c_{,\bar U}$, where the function
$c(\bar U,\bar\theta)$ enters, for example, the expansion of
$g_{\bar\phi\bar\phi}= \bar r^2\sin^2\bar\theta+2c\bar r +O(1)$.
Starting from our metric (3.2), and using the transformation (3.8)
with the functions $\pi,\sigma,\tau, q$ found above, we obtain $c=0$.
Hence the {\it news function vanishes}. In fact, this could be
anticipated since $\psi\sim r^{-2}$ at $I$ ---we are here in the
region in which the tails of cylindrical pulses decay, and there is no
radiation field at null infinity \cite{[12]}. Thus, in these
space-times, the radiation field is all focussed in the direction of
the two ``singular generators'' of $\I$ singled out by the axis (or,
the $\d/\d z$-Killing field). Along these generators, the
Bondi-Penrose radiation field diverges and asymptotic flatness is
lost. In other directions, there is smooth curvature but no flux of
energy.

We conclude this sub-section with a remark.  In their analysis of
isometries compatible with gravitational radiation, Bi\v c\'ak and
Schmidt \cite{[12]} consider axi-symmetric space-times, assume Bondi's
expansion for all $\phi\in [0,2\pi)$ and $\theta\in(\theta_0,
\theta_1)$ and conclude that cylindrical symmetry is not permissible. 
This assertion may seem to contradict the conclusion we just reached
for the time symmetric Einstein-Rosen waves. Note, however, that the
interval of permitted $\theta$'s in the assertion of \cite{[12]}
contains $\theta={\pi \over 2}$, i.e., the directions perpendicular to
the axis of symmetry, while in the present case, Bondi's expansion
fails to hold in that direction. Thus, there is in fact no
contradiction.  In fact, the results obtained in the present work are
fully compatible with those of \cite{[12]}; Bi\v c\'ak and Schmidt
conclude below their Eq.  (52) that, if the function $c=$ vanishes,
the second Killing vector field (in addition to the axial one) can
generate either a time translation or the translation along the axis
of rotation.
\end{subsection}
\goodbreak

\begin{subsection}{Case when ${J\not= 0, L=0}$} 

In this case, Eq. (3.3) tells us that the leading order behavior of
$\psi$ is different: one obtains $\psi\sim{J/{r\cos\theta}}$.
Consequently, transformation (3.8) does not now lead to a Bondi
system; in particular, it does not remove the ``offending'' term in
$g_{rr}\sim r^{-1}$.  Nevertheless, since the leading term in the
metric does not depend on time and is $O(r^{-1})$, typical for static
Weyl metrics, we can attempt to find the required Bondi system by
mimicking the procedure adopted in \cite{[9]}. Let us assume a
transformation of the form
\ba
U &=&\bar U +\pi( r, \bar\theta)\ ,\nonumber \\ 
\theta &=& \bar\theta + \tau^1(\bar\theta)  r^{-1} +\dots\ .
\label{(3.9)}\ea 
Keeping then just the first term in $\psi$ in the expansion (3.3) with
$L=0$, and writing the asymptotic form of the metric analogously to
(3.5), we find that the crucial term $\sim r ^{-1}$ in $\bar g_{rr}$
will vanish if
\be -(\pi_{,r})^2-2\pi_{,r}
+{4J\over{r}}\ \cos\theta=0\ .\label{(3.10)}\ee
Solving in the leading order for $\pi$, we obtain
\be \pi(r, \bar\theta)=2J\cos\bar\theta \ln r + \dots\ .
\label{(3.11)}\ee
In this way we can achieve at least $\bar g_{rr}\sim O(r^{-2})$.
However, with the transformation (3.9) there is no way to satisfy the
requirement $\bar g_{r\bar\theta}=O(1)$. We must admit a logarithmic
term also in the transformation of $\theta$ which, in turn, requires
another logarithmic term in the transformation of $U$.  By assuming
expansions in $r^{-j}\ln^ir$, we find, after some effort, that a
suitable transformation reads
\ba
 U &=&\bar U+\left({2J\over\cos\theta}\cos^2\bar\theta\right)\ln r -
\left(2J^2\sin^2\bar\theta\right){\ln^2r\over r}\, , \nonumber\\
\theta &=& \bar\theta + \left({{2J}{\sin\bar\theta}}\right)
{\ln r\over r} \, . \label{(3.12)}\ea
(Notice that in the leading order $(2J/\cos\theta)\cos^2\bar
\theta=2J\cos\bar\theta$, in agreement with (3.11).)

Now, transforming the metric (3.2), with $\psi$ and $\gamma$ given by
(3.3), (3.4) (with $J\ne 0, L=0$), via (3.12), we obtain the metric in
the following form:
\ba
ds^2 &=-&\left[1-{{2J\over{\cos\bar\theta}}}\ {1\over{r}}+ 
O\left({{\ln r}\over{r^2}}\right)\right]d\bar U^2 \nonumber\\
&-&\left[1-{{2J\sin^2\bar\theta\over{\cos\bar\theta}}}\ {1\over{r}}+
O\left({{\ln^2 r}\over{r^2}}\right)\right]2d\bar Udr \nonumber\\
&+&\left[{4J{\ln r\over r}}+O\left({{\ln^2 r}\over{r^2}}\right) \right]
\ r\sin\bar\theta\ d\bar Ud\bar\theta \nonumber\\ 
&+&\left[ O\left({1\over r^2}\right)\right] dr^2 -
\left[4J{{1\over r}}+O\left({{\ln^2 r}\over{r^2}}\right)\right]r
\sin\bar\theta dr d\bar\theta \nonumber\\
&+&\left[1+ O\left({{\ln^2r}\over{r}}\right) \right]r^2d\bar\theta^2
+\left[1+O\left({{\ln^2 r}\over{r}}\right)\right]r^2\sin^2\bar\theta 
d\phi^2\ . \label{(3.13)}\ea  

Bondi et al \cite{[9]} applied a similar procedure to the Weyl
metrics.  In contrast to their result, however, we did not quite
succeed in bringing our metric to the standard Bondi form. The reason
is that, unlike the Weyl metric, in our case, the leading ``offending''
terms --proportional to $r^{-1}$-- are $\theta$-dependent.  [In the
case of the transformation of the Weyl metric to Bondi's form
--cf. \cite{[9]}-- we have $\pi=2m\ln\bar r+\dots$, $m$ being the
mass. Assuming $m=m(\theta)$ in the Weyl metric ( and thus violating
the field equations), one can make sure that $I$ still exists but the
space-time is only "logarithmically" asymptotically flat.]  By
introducing $\tilde l=r^{-1}$, $\tilde U=\bar U$,
$\tilde\theta=\theta$, $\tilde\phi=\phi$, and rescaling the metric
(3.13) by the conformal factor $\Omega=\tilde l$, we obtain
\ba
d\tilde s^2=\Omega^2ds^2 &=-& \left[1-{{2J}\over{ \cos\tilde\theta}}
\, \tilde l+O(\tilde l^2\ln\tilde l)
\right]\tilde l^2d\tilde U^2 \nonumber\\
&+&\left[1-{{2J\sin^2\tilde\theta}\over{\cos\tilde\theta}}\tilde
l+O(\tilde l^2\ln^2\tilde l) \right] 2d\tilde Ud\tilde l \nonumber\\
&+&\left[-4J(\sin\tilde\theta)\ \tilde l \ln\tilde l+O(\tilde
l^2\ln^2\tilde l)\right]\tilde ld\tilde Ud\tilde l +O(1)\ d\tilde l^2
\nonumber\\
&+&\left[4J\sin\tilde\theta +O(\tilde l\ln^2\tilde l)\right]d\tilde
ld\tilde\theta \nonumber\\
&+&\left[1+O(\tilde l\ln^2\tilde l)\right]d\tilde\theta^2 
+\left[1+O(\tilde l\ln^2\tilde l)\right]\sin^2\tilde\theta d
\tilde\phi^2\ .  \label{(3.14)}\ea 

Thus, the metric is well--behaved as $\tilde l\to 0$, i.e., at $\tilde
l=0$ $\I$ does exist. The metric is continuous on $\I$. However, it is
not differentiable. Thus, it appears that there is a key difference in
the asymptotic behavior in the time-symmetric case and in the general
case. In the general case, $\I$ appears to have a ``logarithmic
character'' \cite{[3],[4]}.  (A word of caution is in order: It is
possible that the differentiability can be improved by continuing the
transformation (3.12) into higher-order terms.)

To conclude, we wish to point out that, although we obtained the
asymptotic forms for $\psi$ and $\gamma$ (Eqs. (3.3) and (3.4))
assuming that the waves have Cauchy data of compact
$(\rho,\phi)$-support, the forms themselves hold in more general cases
as well. An interesting example is provided by the
Weber-Wheeler-Bonnor time--symmetric pulse solution \cite{[13],[14]}.
[The pulse is formed by a linear superposition of monochromatic waves
with a cut--off in the frequency space : $\psi(t,r)=2C\int_0^\infty
e^{-a\omega}J_0(\omega\rho)\cos\omega t d\omega $, where $J_0$ is the
Bessel function and the constant $a$ is an approximate measure of the
width of the pulse. It appears \cite{[17]} that no other integral
containing the Bessel function can be expressed in a closed form,
which apparently makes the Weber--Wheeler--Bonnor pulse "unique" among
non--singular pulse--type solutions of the wave equation in (2+1)
dimensions.]  In this case, we have
\be \psi=\sqrt2C
{\left\{ {\left[(a^2+\rho^2-t^2)^2+4a^2t^2\right]^{1/2}
+a^2+\rho^2-t^2}\over {(a^2+\rho^2-t^2)^2+4a^2t^2}
\right\}^{1/2}}\ ,\ \ a= {\rm const}
\label{(3.15)}\ee
and
\be {\gamma={{1\over2}}C^2}\left\{{1\over{a^2}}-{{
2\rho^2\left[(a^2+\rho^2-t^2)^2-4a^2t^2\right]\over
{\left[(a^2+\rho^2-t^2)^2+4a^2t^2\right]^2}}\ 
+{{1\over a^2}}{{\rho^2-a^2-t^2}\over{
\left[(a^2+\rho^2-t^2)^2+4a^2t^2\right]^{1/2}}}
}\right\}\ .\label{(3.16)}\ee
At $t=0$, the Cauchy data for $\psi$ are $\psi=C(a^2+\rho^2)^{-{1\over
2}}$ and $\psi_{,t}=0$. Nevertheless, expressing the asymptotic forms
of $\psi$ and $\gamma$ at $U=t-r=const$, $\theta=const$, $\phi=const$,
we find after somewhat lengthy calculations (or by using MATHEMATICA),
that $\psi$ and $\gamma$ have asymptotically {\it exactly} the form
(3.3), (3.4) with $J=0$ and $L=-2Ca$.  Therefore, in the directions
not perpendicular to the symmetry axis, these waves do admit a smooth
$\I$.

Similarly, the asymptotic forms of $\psi$ and $\gamma$ with $J\ne 0$
may hold even though the Cauchy data are not of compact support. A
simple prototype, discussed by Carmeli \cite{[15]}, for example, has
\be
\psi={1\over {2\pi}}{{f_0}\over {\sqrt{t^2-\rho^2}}} \ ,\ \ \ 
\gamma={1\over {8\pi^2}}\ {{f_0^2\rho^2}\over {{(t^2-\rho^2)}}} 
\ ,\ \  f_0={\rm const}\ .
\label{(3.17)}\ee

This wave is singular at $t^2=\rho^2$ but it represents the late time
behavior of the solution given by
\be \psi={1\over{2\pi}}\ \int_{-\infty}^\tau
{{f(t')dt'}\over{[(t-t')^2-\rho^2]^{1/2}}}\ , 
\ \ \tau=t-\rho\ , \label{(3.18)}\ee
where $f(t)\ne 0$ only for $0<t<T$; $f_0=\int_0^T\!\!f(t')dt'$. With
this wave we find $\psi$ and $\gamma$ to behave (at $U=t-r=const$,
$\theta=const$, $\phi=const$, $r\to\infty$) according to (3.3) and
(3.4) with $J={{f_0}/{2\pi}}$. The fall--off is now slower but a
``logarithmic'' null infinity, $\I$, does exist.
\end{subsection}
\end{section}
\bigskip\bigskip
\goodbreak
{\bf Acknowledgements:} AA and JB would like to thank the
Albert--Einstein--Institut for its kind hospitality. AA was supported
in part by the NSF grants 93-96246 and 95-14240 and by the Eberly
Research Fund of Penn State University. JB was supported in part by
the grants GACR--202/96/0206 and GAUK--230/1996 of the Czech Republic
and the Charles University, and by the US-Czech Science and Technology
grant 92067.
\bigskip\bigskip\goodbreak

\appendix
\begin{section}{Appendix: Relation between Riemann tensors 
in 3 and 4 dimensions}
\label{A}

The Einstein--Rosen metric (2.1) in coordinates $x^0=u=t-\rho$,
$x^1=\rho$, $x^2=\phi$, $x^3=z$ becomes 
\be ds^2=e^{2(\gamma-\psi)}(-du^2-2dud\rho)+\rho^2e^{-2\psi}d\phi^2
+e^{2\psi}dz^2\ .\label{A1}\ee
Assuming the expansion (2.10) for $\psi$, we know that $\gamma$ can be
written in the form (2.23), and in principle the Riemann (Weyl) tensor
of the vacuum (3+1)--dimensional space-time and its asymptotic
behavior can be obtained from (A1).

However, it is possible to use directly the ``reduction formulas'' for
the calculation of the Riemann tensor of spaces which admit an abelian
isometry group \cite{[16]}. In the coordinates $(x^\mu)=(u, \rho,
\phi, z)$ in which the Killing trajectories orthogonal to the
hyper-surfaces $z=const$ are just $u=const, \rho=const, \phi=const$,
the 4--dimensional components of the Riemann tensor are given by the
following relations (see Eq. (2.3.4) of \cite{[16]} where, however,
the Riemann tensor with the opposite sign is used):
$$ {}^{(4)}R_{abcd}= \bar R_{abcd}, \ \ \ {}^{(4)}R_{3abc}=0\ , $$ 

\be {}^{(4)}R_{3a3b}= -VV_{||ab},\ \ \ V=e^\psi \ ,\label{A2}\ee
where the $"|"$ denotes the covariant derivative with respect to the
metric $\bar g_{ab}$ given by (A1) with $z=const$. The covariant
derivatives $V_{\parallel ab}$ are given in terms of the Christoffel
symbols listed by Stachel \cite{[2]} in his Eq.  (A1) in Appendix in
$(u,\rho,\phi,z)$ coordinates.  [{In Stachel's list of $\Gamma$'s the
following symbols are missing: $\Gamma^{\phi}_{\rho\phi}=
\rho^{-1}-\psi_{,\rho}$, $\Gamma^{\rho}_{\rho\rho}=
2(\gamma_{,\rho}-\psi_{,\rho})$, $\Gamma^z_{\rho z}=\psi_{,\rho}$.
Notice also that his $x^2=z$, $x^3=\phi$, while here we put
$x^3=z$. He treats waves with both polarizations so we must put his
$\chi=0$ when comparing his results with ours.] Using these,
$$ {}^{(4)}R_{3030}= -e^{2\psi}\left[
\psi_{,uu}+3\psi_{,u}^2-2\psi_{,u}\psi_{,\rho}+\psi_{,\rho}^2 
+\gamma_{,\rho}(\psi_{,u}-\psi_{,\rho})+\gamma_{,u}
(\psi_{,\rho}-2\psi_{,u})\right] \ , $$
$$ {}^{(4)}R_{3131}= -e^{2\psi}\left[
\psi_{,\rho\rho}+ 3\psi_{,\rho}^2 -2\gamma_{,\rho}
\psi_{,\rho}\right] \ , $$
$$ {}^{(4)}R_{3232}= -e^{2\psi-2\gamma}\rho^2\left[
2\psi_{,u}\psi_{,\rho}-\psi_{,\rho}^2 
+\rho^{-1}(\psi_{,\rho}-\psi_{,u})\right] \ , $$
\be {}^{(4)}R_{3031}= -e^{2\psi}\left[
\psi_{,u\rho}+ \psi_{,u}\psi_{,\rho} -\gamma_{,\rho}
\psi_{,\rho}+\psi_{,\rho}^2\right] \ . \label{A3}\ee
These are the non-vanishing components ${}^{(4)}R_{3a3b}$ in the
coordinates $(u,\rho,\phi,z)$.  Transforming them back to the
coordinates $ (t,\rho,\phi,z)$ we find --- after projecting them on
the orthonormal tetrad used by Stachel --- precisely his components
$R_{0202}$, $R_{1212}$, $R_{2323}$ and $R_{2021}$ given in his
equations (A3). (They have the opposite signs because Stachel uses the
signature $+ - - - $.)

The components $\bar R_{abcd}$ (formed from the metric $\bar g_{ab}$)
can be expressed in terms of our (2+1)--dimensional Riemann tensor
given in Appendix A of \cite{[1]}. This is formed from the 3--metric
$g_{ab}=e^{2\psi}\bar g_{ab}$; hence we use the behavior of the
3--dimensional Riemann tensor under conformal rescalings (see
e.g.\cite{[16]}, Eq. (2.4.6)).  We find
\be \bar R_{abjk}=2e^{2\psi} R_{abjk}- {1\over 2}e^{-4\psi}
(g_{j[a}V_{b]k}-g_{k[a}V_{b]j})\ ,
\label{A4}\ee
where
\be V_{ik}=2e^{2\psi}\left[(g^{lm}\psi_{,l}\psi_{,m})
g_{ik}-2\psi_{,i} \psi_{,k}-2\psi_{;ik}\right]\ . 
\label{A5}\ee
Here the semicolon denotes the covariant derivative with respect to
the 3--metric $g_{ab}$ (see Appendix A of \cite{[1]} for the
Christoffel symbols).  The non-vanishing quantities $V_{ik}$ turn out
to be the following:
$$ V_{00}=2e^{2\psi}\left[2\psi_{,u}\psi_{,\rho}-
\psi_{,\rho}^2-2\psi_{,u}^2 -2\psi_{,uu} +2\psi_{,u}
(2\gamma_{,u}-\gamma_{,\rho})+2\psi_{,\rho}
(\gamma_{,\rho}-\gamma_{,u})\right]\, $$
$$ V_{01}=V_{10}= 2e^{2\psi}\left[ -\psi_{,\rho}^2-2
\psi_{,u\rho}+2\gamma_{,\rho}\psi_{,\rho}\right]\ ,$$ 
$$  V_{11}= 4e^{2\psi}\left[-\psi_{,\rho\rho}-\psi_{,\rho}^2+
2\gamma_{,\rho}\psi_{,\rho}\right]\ , $$ 
\be V_{22}= 2e^{2\psi-2\gamma}\rho\left[\rho(\psi_{,\rho}^2
-2\psi_{,\rho}\psi_{,u})+2(\psi_{,u}-\psi_{,\rho})\right]\ .
\label{A6}\ee
By substituting these expressions into (3.3) and using the components
$R_{abcd}$ from Appendix A of \cite{[1]}, we find $\bar R_{abjk}$ ---
and thus also ${}^{(4)}R_{abjk}$ --- in the coordinates
$(u,\rho,\phi,z)$. By transforming them to $(t,\rho,\phi,z)$ we
exactly recover Stachel's expressions given in his Eq. (A.3). }
\end{section} 

\bigskip\bigskip\goodbreak


\begin{thebibliography}{99}

\bigskip
\bibitem{[1]} A. Ashtekar, J. Bi\v c\'ak and B. G. Schmidt, ``Asymptotic
Structure of Symmetry Reduced General Relativity" (preceding paper).

\bibitem{[2]} J. Stachel, J. Math. Phys. {\bf 7}, 1321 (1966).

\bibitem{[3]}  J. Winicour, Found. Phys. {\bf 15}, 605 (1985).

\bibitem{[4]} L. Andersson and P. T. Chru\'sciel, Phys. Rev. Lett. 
{\bf 70}, 2829 (1993); P. T. Chru\'sciel, M. MacCallum, and D.
Singleton, Phil. Trans. R. Soc. London {\bf 350}, 113 (1995).

\bibitem{[5]} J. Bi\v c\'ak and B. G. Schmidt, Phys. Rev. D {\bf 40}, 
1827 (1989).

\bibitem{[6]} D. Kramer, H. Stephani, E. Herlt, and M. MacCallum, 
{\it Exact solutions of Einstein's field equations} (Cambridge
University Press, Cambridge, England 1980).

\bibitem{[16]} P. Jordan, J. Ehlers, and W. Kundt,  Akad. Wiss. Lit. 
Mainz. Abhandl., Math.--Naturwiss. Klasse, Nr 2 (1960). 

\bibitem{[7]} G. B. Whitham, {\it Linear and nonlinear waves} 
(J. Wiley, New York 1974). 

\bibitem{[8]} R. Penrose, Phys. Rev. Lett. {\bf 10}, 66 (1963); 
Proc. R. Soc. London {\bf A284}, 159 (1965).

\bibitem{[9]} H. Bondi, M. van der Burg, and A. Metzner, 
Proc. R. Soc. London {\bf A269}, 21 (1962).

\bibitem{[10]} J. Bi\v c\'ak,  Proc. R. Soc. London  {\bf A302}, 201 
(1968). 

\bibitem{[11]} J. Bi\v c\'ak and B. G. Schmidt, Class. Quantum Grav. 
{\bf 6}, 1547 (1989).

\bibitem{[12]} J. Bi\v c\'ak and B. G. Schmidt, J. Math.Phys. 
{\bf 25}, 600 (1984).

\bibitem{[13]} J. Weber and J. Wheeler, Rev. Mod. Phys. {\bf 29}, 509 
(1957).

\bibitem{[14]} W. B. Bonnor, J. of Math. and Mech. {\bf 6}, 203 (1957).

\bibitem{[17]} I. S. Gradshteyn and I. M. Ryzhik, {\it Table of
Integrals, Series, and Products} (Academic Press, New York 1980).

\bibitem{[15]} M. Carmeli, {\it Classical fields: general relativity 
and gauge theory} (J. Wiley, New York 1982).


\end{thebibliography}
\end{document}